\documentclass[a4paper,12pt]{article}
\usepackage{helvet}
\usepackage{amsthm}
\newtheorem{theorem}{Theorem}
\title{A New Structural Property of SAT}
\author{Silvano Di Zenzo \\ Department of Computer Science, University of Rome}
\date{curbastro@tiscali.it}
\begin{document}
\maketitle

\begin{abstract}
We review a minimum set of notions from our previous paper on structural properties of $SAT$ \cite{dizenzoarxiv} that will allow us to define and discuss the ``complete internal independence" of a decision problem. This property is strictly stronger than the independence property that was called ``strong internal independence" in \cite{dizenzoarxiv}.

Next, we show that $SAT$ exhibits this property. We argue that this form of independence of a decision problem is the strongest possible for a problem.

By relying upon this maximally strong form of internal independence, we reformulate in more strict terms the informal remarks on the exponentiality of $SAT$ that concluded our previous paper \cite{dizenzoarxiv}. The net result of that reformulation is a hint for a proof for $SAT$ being exponential. We conjecture that a full fledged proof of that proposition can be obtained by strictly following the line of that hint of proof.
	
\end{abstract}

\section{Introduction}
In a previous paper \cite{dizenzoarxiv} we used strings (borrowed from Computability \cite{odifreddi}) in order to formulate various notions of internal independence of a decision problem. We proved that $SAT$ exhibits a structural property that we called ``strong internal independence." We could then prove various results on $SAT$ using that property.
	
In this paper we prove that SAT exhibits a strictly stronger form of structural independence, that we call ``complete internal independence." We argue that this form of independence of a decision problem is the strongest possible for a problem. Using this stronger property, we will be able to give more explicit form to the informal remarks on the time complexity of $SAT$ that concluded our previous paper. As was the case for that paper, these remarks are only intended to suggest that there is relation between the structure of the kernel of a decision program $P$ (as defined in cited paper) and the time complexity of $P$.
	
To prove this stronger property of $SAT$ we do not need the full machinery of concepts and techniques that were developed in \cite{dizenzoarxiv} (these were used to prove that $SAT$ has no wizards, and that all programs that solve SAT have same kernel). So, we shall take advantage of our present simplified context, and will offer a development of the most elementary aspects of the theory at an introductory level. We shall also take occasion to put in due evidence the roots of the theory: We take ideas from previous work of Scott \cite{scott}, and Larsen and Winskel \cite{larsen}, and use results of Di Zenzo, Bottoni, Mussio \cite{dizenzo}.

\section{Introductory Remarks on Strings}
We define a string to be a partial function $f : N \rightarrow \Sigma$ having finite domain. $N$ is the set of the positive integers and $\Sigma$ is an alphabet which includes $0$ and $1$. In simple words, a string $g$ being included (or subsumed) in a word $x$ is that which remains of that word if we cancel out zero or more letters, while leaving blanks in places of letters. We define $\Sigma_\infty$  to be the set of all strings over $\Sigma$.
	
We regard a $g \in \Sigma_{\infty}$  as a prescription that a word $x$ over $\Sigma$ may or may not satisfy. If word $x$ includes string $g$, we think that $x$ satisfies prescription $g$. Two strings are $compatible$ if they prescribe same values to arguments in the intersection of their domains. Otherwise they are $incompatible$.
	
If $Dom(f )$ is an initial segment of $N$, $Dom(f )=\{1,.., n\}$ for some positive integer $n$, then we say that $f$ is a full string or word of length $n$ over $\Sigma$. Thus, words are certain special strings. It is desirable to have a reserved name for describing the new mathematical objects, and we consider that ``string" is the appropriate one: It is required that one remembers that ``string" and ``word" have different meanings.

The new objects, the strings, can be viewed as generalized words which can undergo splittings and rejoinings according with more flexible schemes as compared with words. The strings actually occur as generalized words: They represent fragments of words, and we think of them as partial words.

There is a duality between strings and words, and one of the challenges of this theory is to formalize this duality in a useful manner. This is the purpose of the Galois connection in \cite{dizenzoarxiv}. In this paper we do not need the full machinery of that connection, and we will just use some of the results. The important result is that whenever we have a set $A$ of words, we can set up a second set $Log(A)$ (or $Log_E(A)$ in a relativized theory) that one may wish to regard as $jinnee$ or $genie$ of given set $A$. The new set, the $logogram$ of $A$, is a set of strings: the $genes$ (so to speak)  of the words in $A$.
		
When we deal with strings we think of a game in which one reconstructs words from strings. Equivalently, the game is to classify a word by matching it with strings taken from some vaste repository of strings. There are various basic operations that we can perform on strings. These include (i) to join strings to form new strings and (ii) to extend a string to form a word. These seem to be the very basic operations on strings.

If we join two incompatible strings we just get the void set $\emptyset$, which is to say that output vanishes. If we join two compatible strings we obtain a new string which encapsulates each of the given strings.
	
The join of two compatible strings $f, g$ is a new string that we denote $f+g$. In order that the operation of join may be applicable to any pair of strings, we shall also define the join of $g$ and $g$. This we take to be $g$ itself. Thus join is idempotent.
	
Further, in order to be able to iterate the operation of join we must define joins of sets of strings. This is an important step in our theory, and yields surprisingly reach algebraic structures. We define the join of two nonempty sets of strings $H, K$ to be the set of all strings formed by joining one string from $H$ with one string from $K$ and aggregating all the joins formed in this way in a new set of strings $H+K$. The move of focus from join of single strings to that of whole sets of strings opens wide horizons.

\section{Introductory Remarks on Independence}
In this paper we make extensive use of the strings and of their associated sets, the cylinder sets. As usual in various fields of application, it can be convenient for us to speak of special classes of sets (the cylinder sets, in this case) as being events: This will also make the recourse to arguments using notions of dependence/independence as natural as possible. In a sense, it suggests these arguments.
	
To every event $A$ there corresponds a contrary event ``not $A$," to be denoted $A^c$. Event $A^c$ occurs if and only if $A$ does not occur. Always remember that we are considering words and sets of words, and the whole set is $\Sigma^*$.
	An event may imply another event: $A$ implies $B$ if, when $A$ occurs, then $B$ necessarily occurs. If $A$ implies $B$ and also $B$ implies $A$, then we conclude that $A$ and $B$ are one and the same event. Events are combined into new events by means of operations expressed by $and$, $or$ and $not$. The event ``$A$ and $B$" is denoted $A \cap B$, or simply $AB$. It occurs if and only if both the event $A$ and the event $B$ occur. $AB$ cannot occur if $AB = \emptyset$. This is the case when, if $A$ occurs then $B$ does not occur, and if $B$ occurs then $A$ does not occur. When $AB = \emptyset$  we say that $A$ and $B$ are $incompatible$.
    Event ``$A$ or $B$" is denoted $A \cup B$.

Let us have a more detailed look at incompatibility between events. Assume that both $A$ and $B$ are nonempty and incompatible. Then $AB = \emptyset$. Then we can rely upon the following three circumstances (i) That there is a word $x$ which is in $A$ but not in $B$, (ii) That there is a word $y$ which is in $B$ but not in $A$, (iii) That there are no words $z$ being both in $A$ and in $B$.
      It is not guaranteed that there exists a word $w$ which is neither in $A$ nor in $B$. For this fourth circumstance to hold the additional requirement that $A \cup B$ shall not be coincident with the totality of all words $\Sigma^*$ should be satisfied.

\section{Introductory remarks on NP relations}
The applications that we have on agenda deal with certain structural properties that a decision problem can exhibit. When present, these properties express internal independence of the problem. To correctly formulate these properties, we need to refer to a decision problem as a pair $(E, F)$ of sets, the second included in the first. Indeed, the properties that we establish are about $E, F$, and how $F$ is embedded in $E$.

Our main objective consists of the decision applications, in particular the application to $SAT$. In this paper (as in companion paper \cite{dizenzoarxiv}) we set forth new techniques to attack the decision problems. We consider a generic decision problem $\Pi$  and assume that the instances of $\Pi$ are encoded as words over some fixed problem alphabet $\Sigma$. We call $E$ the set of those words that encode instances of $\Pi$. Obviously, $E$ is a recursive set, and we assume it is infinite. In any decision problem  $\Pi$ there is a second recursive set $F$ being a proper subset of $E$. The words in $F$ encode those instances of $\Pi$  that we intend to recognize from all of the remaining instances. We refer to this encoded decision problem as problem $(E, F)$.

Thus, we define an (encoded) decision problem to be a pair $(E, F)$ of recursive sets of words over an alphabet $\Sigma$ where $F$, the target set of the decision, is a subset of the reference set $E$. For example, in the problem of deciding satisfiability of boolean formulas, $E=CNF$ is the set of the strings that encode formulas in conjunctive normal form, and $F=SAT$ is the subset of the satisfiable formulas.

\paragraph{Relations over an alphabet} Let $G$ be a subset of the Cartesian product $\Sigma^*\times\Sigma^*$ so that $G$ is a relation on words over $\Sigma$. There exist exactly one set $A$ and exactly one set $B$ with the properties (i) $x \in A$ is equivalent to relation $\exists y G(x, y)$, (ii) $y \in B$ is equivalent to relation  $\exists x G(x, y)$. These sets are first and second projection of relation $G$. The first projection will be denoted $Dom(G)$, the second is $Cod(G)$.

\paragraph{NP relations} We say $G$ is decidable in polynomial time if there exists a deterministic algorithm which decides membership in $G$ of any pair $(x, y)$ in time polynomial in the length $|x|$ of the input instance $x$. We say that $G$ is polynomially balanced or else polynomially bounded if there exists a polynomial $p$ such that $(x, y)$ in $G$ implies $|y| \le p(|x|)$.

A relation $G$ which is both polynomial-time decidable and polynomially balanced is an NP relation \cite{papa}. A language $L$ is in NP iff there exists an NP relation $G$ such that $L=Dom(G)$. In short, $L$ is in NP iff $L$ is the first projection of some NP relation. When this happens, we say $G$ is a defining relation for $L$. Note that a language $L$ being in NP may have more than one defining relations, which is to say $L$ can be first projection of more than one NP relations.

With each NP relation $G \subseteq \Sigma^*\times\Sigma^*$ we associate the following search problem: Given $x$ find $y$ such that $G(x, y)$ or state that no such $y$ exists.

\paragraph{P relations}
Let $G$ be an NP relation. Thus $G$ is polynomially bounded, and there is a search problem associated with $G$. We say that $G$ is a P relation iff there exists a polynomial-time algorithm that given $x$ finds $y$ such that $G(x, y)$ or states that no such $y$ exists. Equivalently, we say that $G$ is a P relation if $G$ is an NP relation and, besides, the search problem associated with $G$ is solvable in polynomial time.
Let $L$ be a language in NP. Then there exists at least one NP relation $G$ which defines $L$. Language $L$ is in P if and only if at least one of the NP relations whose first projection is $L$ is actually a P relation.

\paragraph{Standard characterization of NP}
Let $\Pi$ be a problem in NP, encoded as a pair $(E, F)$ over some problem alphabet $\Sigma$. In the standard characterization of class NP, there exists a sequence $y_1, y_2,..$ of specials words that are called the solutions of problem $(E, F)$. In general, given any generic problem instance $x$ in $E$, we have that $x$ is ``satisfied" by certain solutions. There are also instances $x$ which fail to be satisfied by any solutions, and we call them unsatisfiable.
    What ``satisfaction" means operationally is proper of the problem $(E, F)$ under study.

With any particular NP problem $(E, F)$ one associates a recursive function $\alpha(n)$ such that the solutions that can possibly satisfy an instance $x$ such that $|x|=n$ are all comprised between $y_1$ and $y_{\alpha(n)}$.

Associated with solutions $y_1$, $y_2$,.. there is a decomposition of target set $F$ into subsets $F_i$ called $solution$ $regions$, where $F_i$ is the set of those $x's$ that are satisfied by $y_i$. The obvious relation $F=\bigcup_i F_i$ holds.

\section{Alphabets, strings, and words}
By an alphabet $\Sigma$ we mean a finite set of elements called symbols. Any finite sequence of symbols from $\Sigma$ is called a word over $\Sigma$. The size (or length) of a word $w$, noted $|w|$, is the number of symbols composing the word. The size of a string equals the maximum number in its domain (strings are partial functions, hence every string has a domain). For strings that are words the two numbers coincide.

In this paper ``string" and ``word" are not synonimous. We borrowed strings from Computability, where a string over $\Sigma$ is a partial function $g : N \rightarrow \{0, 1\}$ with finite domain. If the domain of string $g$ is an initial segment of $N$ then $g$ is a word.
	
We write $\Sigma^n$ for the set of all words of length exactly $n$ over $\Sigma$ while $\Sigma^*$ is the set of all words over $\Sigma$ as usual, thus $\Sigma^*$ =  $\bigcup_n\Sigma^n$.

\paragraph{Echelons of strings}
We define $\Sigma_n$ to be the set of all strings of size at most $n$ over the alphabet $\Sigma$. We say that $\Sigma_n$ is an echelon of strings over  $\Sigma$.

The empty string $\bot$ is a member of the set $\Sigma_\infty$ of all strings over $\Sigma$. Its length is 0. The empty string $\bot$ belongs to all echelons of strings. The meet of any two strings in $\Sigma_n$ is a string in $\Sigma_n$. The join of any two compatible strings in $\Sigma_n$ is a string in $\Sigma_n$.

\paragraph{Finite character of languages}
 Let $S$ be any set and $\mathcal E$ any class of subsets of $S$. We say that $\mathcal E$ (as well as its members) are of finite character if there exists a class $\mathcal F$ of finite subsets of $S$ such that any $A \in \mathcal E$ is univocally determined from its intersections $T \cap A$ with the elements $T$ of class $\mathcal F$.

We take the class $\mathcal L = \mathcal L(\Sigma)$ of all languages over $\Sigma$ as class $\mathcal E$ and set $\mathcal F=\{\Sigma^n : n \in N\}$. We then have that any language $L\in \mathcal L$ is univocally determined by its intersections $\Sigma^n \cap L$ with the members of class $\mathcal F$ (it is also interesting to note that these intersections are disjoint taken two by two). By way of consequence, $L$ is of finite character.

Note that the union and the intersection of two languages reduces to the union and the intersection of their respective echelons. More specifically, if $A$ and $B$ are sets of words over some alphabet $\Sigma$, then we may well form the union $A \cup B$. However this is actually done echelon by echelon, namely $A\cup B= \bigcup_n (A \cup B)^n= \bigcup_n (A^n \cup B^n)$. Analogously for the intersection $A\cap B= \bigcup_n (A \cap B)^n= \bigcup_n (A^n \cap B^n)$.

It will scarcely be the case that we consider sets encompassing some of the echelons of a language $L$. In our arguments, either all of the echelons are considered simultaneously (when we are reasoning at large, that is in terms of infinite sets) or else only one of the echelons is under focus (when we reason echelon by echelon). We may well put under focus the union of all the echelons, which is the whole of set $L$ (when we go at large) or else we may take one echelon at a time. In many cases going echelon by echelon is convenient, e.g. when we deal with nonuniform circuit classes. In our previous paper \cite{dizenzoarxiv} we considered single echelons in order to derive our results about SAT.

The finitary character of languages will be of help in our study. It will make it possible to break various infinitary statements down to finitary ones. Besides, it will make it possible for us to regard an infinite set of words as the disjoint union of its echelons (which are finite sets). Note that the finite character of languages manifests itself as algebraicity of the closure operators associated with the Galois connection in \cite{dizenzoarxiv}.

\paragraph{Basic properties of strings}
In this paragraph we establish various technical concepts regarding the objects in $\Sigma_\infty$ namely the strings over $\Sigma$.

Set $\Sigma_\infty$  is partially ordered in a straightforward way. Given any pair of strings $f,  g \in \Sigma_\infty$, we say $g$ is an extension of $f$ (written $f \le g$ or $g \ge f$) as soon as $Dom( f ) \subseteq Dom(g)$ and $g$ takes exactly the same values as $f$ in $Dom(f)$. If $f \le g$ and $g \le f$ then $f = g$. When $f \le g$ but not $f \ge g$, we write $f < g$ and say that $g$ is a proper extension of $f$, or, equivalently, $f$ is a proper substring of $g$.
	
We define $\bot$ to be the null partial function $N \rightarrow \Sigma$. Considered as a prescription, $\bot$ is satisfied by all words $x \in \Sigma^*$. Note that $Dom(\bot)= \emptyset$. We say $\bot$ is the void (or null) element of set $\Sigma_\infty$. Any element in $\Sigma_\infty$ is an extension of the bottom element $\bot$. Thus, the ordered structure $(\Sigma_\infty , \le)$ has a least element $\bot$.

In addition to order, $\Sigma_\infty$ is equipped with a relation of compatibility (or consistency). Two elements $f$ and $g$ of $\Sigma_\infty$ are $compatible$ as soon as $f(x)=g(x)$ for $x$ in $Dom(f) \cap Dom(g)$. If we regard $f$ and $g$ as prescriptions, we say that they are compatible as soon as it is not the case that they assign different values to one and the same entry of a word. If $f, g$ are disjoint, which is to say $Dom( f ) \cap Dom(g)=\emptyset$, then $f$ and $g$ are certainly compatible.We call $f,  g$ $incompatible$ as soon as they fail to be compatible.

Given any pair of compatible strings $f, g$, we define their join $f+g$ to be the least string which is an extension of both $f$ and $g$. Thus $f,  g \le f+g$ and $Dom(f+g)$ is exactly $Dom(f) \cup Dom(g)$. Join is the most peculiar operation that we can perform with two strings as operands. The join of two compatible strings $a, b$ is a new string which subsumes both $a$ and $b$. The join of two incompatible strings is the void set of strings.

To be able to iterate the operation of join we define the join of two sets of strings. Thus we define the join of the nonempty sets of strings $H, K$ to be the set of strings formed by joining each string of $H$ with each string of $K$ and aggregating all the joins formed in this way in one set. Thus, the join of $H$ and $K$ is the set of strings $H+K=\{a+b : a\in H, b \in K\}$.

For any arbitrary subset $H$ of $\Sigma_\infty$ we define $H+ \emptyset = \emptyset +H = \emptyset$.

The meet $f \wedge g$ of any two strings is the restriction of $f$ (or $g$) to that portion of the intersection $Dom(f) \cap Dom(g)$ where $f$ and $g$ agree. Thus $f, g \ge f \wedge g$ and $Dom(f \wedge g)$ is included in $Dom(f) \cap Dom(g)$.

\paragraph{Consistent Sets of Strings}
Let $H$ be any subset of the space $\Sigma_\infty$. We say that $H$ is reduced as soon as no string $g \in H$ is properly included in another string $f \in H$. Given any set of strings $H$, we denote by notation $|H|$ the set of all those strings in $H$ that do not properly include other strings in $H$. It is immediately verified that $H$ and $|H|$ are isoexpansive.

A set of strings $H$ is $consistent$ as soon as there exists a word $x$ which includes every string in $H$. Since any word $x$ includes a finite number of distinct strings, all the consistent sets of strings are finite. If word $x$ includes all of the strings in $H$ then we say that $H$ is consistent by virtue of word $x$.
	
It can be shown that a finite set of strings $H$ is consistent as soon as the strings in it are compatible taken two by two. Note that there exist infinite sets of strings whose strings are compatible taken two by two. However these sets of strings are not consistent even though the strings in it are compatible taken two by two. A set of strings is consistent if and only if there exists a word which includes all of the strings in the set.

\section{Generalized Certificates}
In companion paper \cite{dizenzoarxiv} we generalized the notion of certificate of membership of standard theory of NP relations. Our generalization consists in identifying strings as the appropriate mathematical entities suited for representing the certificates of membership.
The idea is as follows: An input word $x$, whose membership in the reference set $E$ has already been ascertained, actually belongs in the target set $F$ as soon as it includes certain special strings that are characteristic of those words of set $E$ that happen to be satisfiable. So to speak, we assume that satisfiability is accompanied by signs: The observable signs of satisfiability. We assume that the signs are interspersed within the object.

	(As the Irish passerby is manifestly the bearer of signs of Irishmanship - otherwise I would't be able to recognize him as an Irishman among hundreds of Englishmen alongstreet in London - so shall the satisfiable formulas carry the signs of satisfiabilty.)

We assume that, for individuals that are words over an alphabet, the signs can only consist of included strings.
Our main object in present research effort is to construct the theory that will provide us with the set of the strings that characterize the satisfiable instances of an NP problem. What we need is the set of the signs. We will call it the $logogram$ of the target set $F$ relative to the base $E$ as it comes up as a collection of logos. This exercise is the mathematical core of companion paper \cite{dizenzoarxiv}.

\section{Connection between Strings and Words}
In this section we offer an overview of the mathematics through which we attempted the construction of the ``set of the signs" in \cite{dizenzoarxiv}.

To every recursive set $E$ over alphabet $\Sigma$ we associate the set $\Sigma_\infty (E)$, which is that subset of $\Sigma_\infty$  which contains all strings that occur in words of the reference set $E$. Thus, we define
\begin{equation}
\Sigma_\infty (E) = \{g \in \Sigma_{\infty} : (\exists x \in E)  g \le x \}.
\end{equation}
This is the set of all those strings $g$ in $\Sigma_\infty$  whose associated cylinder $Exp(g)$ intersects $E$:
\begin{equation}
\Sigma_\infty (E) = \{g \in \Sigma_{\infty} : Exp(g) \cap E \not= \emptyset\}.
\end{equation}
Thus, $\Sigma_\infty (E)$ is the set of those strings in $\Sigma_\infty$  whose associated cylinder contains elements of set $E$. It is understood that $E$ is the set of words over $\Sigma$ that encode instances of some fixed reference computational problem $\Pi$. (Whenever we talk of a reference set $E$ there is an implicit reference to some fixed computational problem $\Pi$ as well as to a program $P$ solving  $\Pi$.)

In \cite{dizenzoarxiv} we have shown that there is a Galois connection (in the original sense given in \cite{birkhoff}) between the subsets of $\Sigma_\infty (E)$ and the subsets of $E$. The connection induces two closure systems: On $\Sigma_\infty (E)$ on one side (conventionally the left) and on $E$ on the other (the right).

The (involutory) isomorphism associated with the connection is a one-one onto correspondence between closed subsets $F \subseteq E$ on one side and closed subsets $H \subseteq \Sigma_\infty (E)$ on the other. A convenient notation for the image of set $F \subseteq E$ under this isomorphism is $Log_E (F)$, and this suggests notation $E^H$ for the image of $H \subseteq \Sigma_\infty (E)$ through the inverse isomorphism. From the general theory of the Galois connection, $E^H$ is then the set of those $x$ in $E$ which include strings from $H$. We then have
\begin{equation}
E^{Log_E F} = F, Log_E (E^H) = H
\end{equation}
holding for closed sets $F \subseteq E$ and $H \subseteq \Sigma_\infty (E)$.

On the right side of the connection, the subsets $F$ of $E$ which happen to be closed are those that are relative cylinders in $E$. We understand that a subset $F$ of $E$ is a relative cylinder in $E$ as soon as there exists $H \subseteq \Sigma_\infty (E)$ such that $F=E^H=Exp_E (H)$.

It is immediately seen that our notion of a cylinder set is an extension of the ordinary notion of a cylinder set as defined e.g. in \cite{odifreddi}. Our notion of a cylinder reduces to the ordinary notion as soon as we confine to strings that are words. Besides, those subsets $F$ of the reference set $E$ which are cylinders in the ordinary sense can easily be shown also to be relative cylinders with respect to $E$.

In present paper we will exhibit an appropriate encoding for SAT in which the reference set $E$ is prefix-free. Everything becomes easier under this encoding: All subsets of $E$ happen to be closed in $E$, and our theory becomes self-contained (there is no longer need to invoke cylindricity of SAT as proved e.g. in \cite{balcazar}).

\paragraph{Absolute cylinders} Given $g \in \Sigma_\infty$, we define the (absolute) cylinder associated with $g$ to be the set
\begin{equation}
Exp(g) = (\Sigma^*)^g = \{s \in \Sigma^* : s \ge g\}
\end{equation}
Thus, $Exp(g)$ is the set of those words over $\Sigma$ which include $g$. $Exp(g)$ is the $absolute$ $expansion$ of string $g$. The two notations $Exp(g)$ and $(\Sigma^*)^g$ are interchangeable (we find it convenient to keep both). We also say that $Exp(g)$ is $elementary$ to mark difference with the nonelementary cylinders to be introduced below. We call $g$ the $signature$ of set $A=Exp(g)$.

Now assume that, as special case, $x$ is a word over $\Sigma$. Then $Exp(x)$ is the set of all words over $\Sigma$ which exhibit prefix $x$. For $x =\bot$ we have $Exp(\bot)=(\Sigma^*)^\bot=\Sigma^*$.
Thus, $\Sigma^*$ is itself an elementary cylinder set: Its signature is $\bot$.

As next step, we define the general cylinder sets (just cylinders, not necessarily elementary). First, we define the $expansion$ of a set of strings: Let $H \subseteq \Sigma_\infty$  be any set of strings over $\Sigma$. We define $Exp(H)$ to be the set
\begin{equation}
Exp(H)=(\Sigma^*)^H=\{x \in \Sigma^* : (\exists g \in H)  x \ge g \}.
\end{equation}
Thus $Exp(H)$ is defined to be the set of all words in $\Sigma^*$ which subsume strings from $H$. That given, a set $A \subseteq \Sigma^*$ is defined to be a $cylinder$ as soon as there is $H \subseteq \Sigma_\infty$ such that $A=Exp(H)$.

Note that the empty set of words is a cylinder since $(\Sigma^*)^{\emptyset} = \emptyset$ and $\emptyset$ is a subset of $\Sigma_\infty$. However, $\emptyset$ is not an elementary cylinder.

The last notion that we give in this list of notions is $cylindrification$. By this, we understand expansion restricted to sets of words: Given $A \subseteq \Sigma^*$, the cylindrification of $A$ is $Exp(A) = (\Sigma^*)^A$. By definition, this is the set of all words in $\Sigma^*$ which include words from $A$. Now a word is a string whose domain consists in an initial segment of $N$. Thus, a word $y$ subsumes another word $x$ if and only if $x$ is a prefix of $y$. Thus, $Exp(A)$ = words in $\Sigma^*$ which exhibit a word from $A$ as prefix.

\begin{theorem}
$Exp$ is a closure operation in $\Sigma^*$. By the way, a topological one.
\end{theorem}
\begin{proof}
(I) Every word in $\Sigma^*$ is a prefix of itself, hence every word in $A$ has a word from $A$ as prefix. 	 Thus, $A \subseteq Exp(A)$.

(II) $Exp(Exp(A))=Exp(A)$. Indeed, if a word $z$ belongs to first member, then $z$ has a prefix $y \in Exp(A)$. Then $y$ has a prefix $x \in A$. We conclude that $x$ is a prefix of $z$, hence $z$ belongs in the second member. 	Equality of the two members follows from $Exp(A) \subseteq Exp(Exp(A))$ by (I).

(III) Let $A \subseteq B \subseteq \Sigma^*$. If any word $x$ has a word from $A$ as prefix then $x$ has a prefix from set $B$ also.
	
This yields $Exp(A) \subseteq Exp(B)$.

This completes the proof that $Exp$ is a closure operation in $\Sigma^*$.

(IV) Let $A, B$ be any pair of sets of words in $\Sigma^*$. Then $Exp(A\cup B)=$ words over $\Sigma$ that either have a prefix from $A$ or from $B$.

Clearly $Exp(A\cup B) = Exp(A) \cup Exp(B)$. Thus, $Exp$ is topological.
\end{proof}

\begin{theorem}
$A \subseteq \Sigma^*$ is a cylinder set iff $Exp(A)=A$.
\end{theorem}
\begin{proof}
(I) Let $A$ be a cylinder set. Then there is $H \subseteq \Sigma_\infty$  such that $A=Exp(H)$. By taking the expansion of both sides we get $Exp(A)=Exp(Exp(H))$.

We show that $Exp(Exp(H))=Exp(H)$.
	
Indeed, let $z$ be a word belonging to first member. Then $z$ includes a word $y \in Exp(H)$. Being a member of $Exp(H)$, $y$ includes a string $g$ from $H$. Then $z$ includes $g$. Since $z$ includes a string from $H$, it belongs to $Exp(H)$.

Thus, $Exp(Exp(H)) \subseteq Exp(H)$.
	
By Theorem 1, $Exp$ restricted to $\Sigma^*$ is a closure operator. Hence $Exp(H) \subseteq Exp(Exp(H))$.

By way of consequence, $Exp(Exp(H))=Exp(H)$.
	
Since $A=Exp(H)$, we conclude that $Exp(A)=A$.

(II) 	Let $Exp(A)=A$.
	
Words are certain special strings. Thus, we may well affirm that there is a set of strings $A$ such that $A=Exp(A)$. Then $A$ is an absolute cylinder.
\end{proof}

\begin{theorem}
(I) The intersection of two cylinder sets is a cylinder set, (II) The union of two cylinder sets is a cylinder set.
\end{theorem}
\begin{proof}
(I) Let $A, B$ be cylinder sets. Since $A$ is a cylinder, there exists $H \subseteq \Sigma_\infty$ such that $A=Exp(H)= (\Sigma^*)^H$. Analogously, there exist $K \subseteq \Sigma_\infty$  such that $B=Exp(K)=(\Sigma^*)^K$.

It follows that $A \cap B=(\Sigma^*)^H \cap (\Sigma^*)^K$.

The latter is the set of all words over alphabet $\Sigma$ each of which includes at least one string from $H$ and at least one from $K$.

Let $x$ be any word over alphabet $\Sigma$. It is easily seen that $x$ includes a string from $H$ and one from $K$ if and only if $x$ includes their join, thus $(\Sigma^*)^H \cap (\Sigma^*)^K=(\Sigma^*)^{H+K}$. Since $(\Sigma^*)^{H+K}$ is a cylinder, we conclude that the intersection of two cylinder sets is a cylinder set.

(II) Let $A=Exp(H)$ and $B=Exp(K)$ be cylinder sets.

Here $H, K \subseteq \Sigma_\infty$.

Then $A \cup B = (\Sigma^*)^H \cup (\Sigma^*)^K$ is the set of all words over $\Sigma$ each of which includes a string from $H \cup K$.

Thus, $A \cup B = (\Sigma^*)^H \cup (\Sigma^*)^K = (\Sigma^*)^{H \cup K}$.

Since $(\Sigma^*)^{H \cup K}$ is a cylinder, we conclude that the union of two cylinder set is a cylinder.
\end{proof}
We conclude this subsection with one further remark on the structure of the absolute cylinders. Let $A$ be an absolute cylinder. From Theorem 2, we have that $A$ happens to be the same set as its own cylindrification. This is to say that $A$ contains, together with any of its words $x$, all of the words that exhibit prefix $x$.

\paragraph{Relative cylinders} Given $H \subseteq \Sigma_\infty (E)$, we define
\begin{equation}
Exp_E (H) = E^H = \{x \in E : (\exists a \in H)  x \ge a \} = E \cap Exp(H).
\end{equation}
Thus, $E^H$ is the set of those words in $E$ which contain strings from $H$. We call $E^H$ the expansion of $H$ relative to base $E$. For $E= \Sigma^*$ we regain the absolute expansion of $H$.

Given any set $A \subseteq E$, we define $A$ to be a $cylinder$ $in$ $E$ as soon as there is $H \subseteq \Sigma_\infty (E)$ such that $A=E^H$.

Explicitly note that $E^H$ is the intersection between an absolute cylinder $Exp(H)$ and the reference set $E$. We actually regard $E^H$ as being a relativized cylinder, i.e., a cylinder relative to an underlying set $E$.

We saw that cylindrification is a closure operation in $\Sigma^*$. Besides, we proved that $A \subseteq \Sigma^*$ is a cylinder set as soon as $Exp(A)=A$. We derived these results in the absolute case. How do these results reformulate when we replace $\Sigma^*$ with an infinite recursive set of words $E$?
\begin{theorem}
Cylindrification $Exp_E$ relative to a reference set $E$ is a closure operation in $E$. By the way, a topological closure.
\end{theorem}
\begin{proof}
(I) Every word in any set of words $E$ is a prefix of itself, hence every word in $A \subseteq E$ is a word of set $E$ having a word from $A$ as prefix. Thus, it belongs to $Exp_E (A)$. Thus, $A \subseteq Exp_E (A)$.
	
(II) $Exp_E (Exp_E (A))=Exp_E (A)$. Indeed, if word $z$ belongs to first member, then $z$ is a word from $E$ having a prefix $y \in Exp_E (A)$. But then $y$ has a prefix $x \in A$. Then $x$ is a prefix of $z$, hence $z$ belongs in the second member. Equality of the two members follows from $Exp_E (A) \subseteq Exp_E (Exp_E(A))$ by (I).
	
(III) Let $A \subseteq B \subseteq E$. If any word $x \in E$ has a word from $A$ as prefix then $x$ has a prefix from set $B$. This yields $Exp_E (A) \subseteq Exp_E (B)$.
	
This completes the proof that $Exp_E$ is a closure operation in $E$.
	
(IV) Let $A, B$ be any pair of sets of words in $E$. Then $Exp_E (A \cup B)$ = words in $E$ that either have a prefix from $A$ or from $B$. Clearly $Exp_E (A \cup B) = Exp_E (A) \cup Exp_E (B)$. Thus, $Exp_E$ is topological.
\end{proof}
\begin{theorem}
$A \subseteq E$ is a cylinder in $E$ iff $Exp_E (A)=A$.
\end{theorem}
\begin{proof}
Similar to that of Theorem 2.
\end{proof}
Thus, Theorems 1 and 2 do hold also in the relative case. By way of consequence, given $A \subseteq E$, we have that $A$ is a cylinder in $E$ if and only if $Exp_E (A) = A$. Thus, $A$ is a cylinder in $E$ if and only if $A$ is closed under cylindrification relative to $E$.

We conclude this subsection with a few remarks on compatibility and relative compatibility of strings. Let $f, g$ be strings in $\Sigma_\infty (E)$. Then there exist in $E$ words which include $f$ as well as words which include $g$. If $f$ and $g$ are incompatible, then $E$ certainly does not have words including both $f$ and $g$. However, even if $f$ and $g$ are compatible, it is not necessarily the case that $E$ shall have to contain words which simultaneously include both $f$ and $g$.
	The whole question can well be reformulated as follows. Let us assume $f$ and $g$ compatible. Then $f+g$ exists, but it is not mandatory that it belongs to $\Sigma_\infty (E)$. If $E$ fails to contain a word including both $f$ and $g$, then $f$ and $g$ are compatible and nevertheless they are inconsistent relative to reference set $E$. When that is the case, the join $f+g$ fails to belong in $\Sigma_\infty (E)$. $f$ and $g$ compatible is necessary for $f+g$ to exist in $\Sigma_\infty (E)$, but the same condition is by no means sufficient to conclude $f+g \in \Sigma_\infty (E)$.

\paragraph{Absolute and relative logograms} In this section we first introduce the logogram of a set of words relative to a reference set $E$. Next, we regain the absolute logogram for $E=\Sigma^*$. For any $F \subseteq E$, we define
\begin{equation}
Log_E (F)=\{g \in \Sigma_\infty (E) : (\forall s \in E) s\ge g \Rightarrow s \in E^F \}
\end{equation}
to be the logogram of set $F$ to base $E$. Note that, since words are strings, $E^F$ is defined. $E^F$ is the relative cylindrification of $F$ in $E$, and this in turn (since $F$ is a set of words) is the set of all words in $E$ that are prefixed by words in $F$. Observe that, if $F$ is a cylinder in $E$, which is to say if $F=E^H$ for some $H \subseteq \Sigma_\infty (E)$, then $E^F=F$ by Theorem 5.

Besides, $Log_E (F)$ is the set of those strings $g$ whose associated cylinder $Exp(g)$ cuts out of $E$ an intersection set $Exp(g) \cap E$ which is (i) nonvoid, (ii) fully contained in the cylindrification $E^F$ of $F$ relative to $E$. For $E=\Sigma^*$, the above defining equation for the logogram is rewritten
\begin{equation}
Log (F)=\{g \in \Sigma_\infty : (\forall s \in \Sigma^*) s\ge g \Rightarrow s \in Exp(F) \}
\end{equation}
This is what we call the absolute logogram of language $F$. Thus, $Log(F)$ is the set of the strings $g$ whose associated absolute cylinder $Exp(g)$ is fully contained in $Exp(F)$.

We conclude this section listing two properties of mapping $Log_E$ that are used in the sequel. First, given $A, B \subseteq E$, one has:
\begin{equation}
A \subseteq B  \Rightarrow  Log_E (A) \subseteq Log_E (B).
\end{equation}
This is monotonicity of the logogram. The second property is:
\begin{equation}\label{mauri8}
Log_E (A \cup B) \supseteq Log_E(A) \cup Log_E(B).
\end{equation}
Let us understand this inclusion. $Log_E (A \cup B)$ is the set of all strings that, for $x$ in $E$, are able to trigger event $x\in E^{A \cup B}=E^A \cup E^B$. A string that triggers $x\in E^A$ certainly belongs to $Log_E (A \cup B)$. Analogously, a string that triggers $x\in E^B$ certainly belongs to $Log_E (A \cup B)$. Thus, $Log_E(A) \cup Log_E(B)$ certainly is a subset of $Log_E (A \cup B)$. However, there can be strings $f$ whose inclusion in a word $x \in E$ is a sufficient condition for event $x\in E^A \cup E^B$ but not for $x\in E^A$ or $x\in E^B$. Thus, in the general case $Log_E (A \cup B)$  is not the same set as $Log_E(A) \cup Log_E(B)$.

\begin{theorem}
Theorem Let $H \subseteq \Sigma_\infty (E)$. The mapping which carries $H$ onto $Log(Exp(H))$ is a closure operator.
\end{theorem}
\begin{proof}
Proof. $Exp(H)$ is the set of all words in $\Sigma^*$ which include strings in $H$. Given any set of words $A \subseteq \Sigma^*$ we have that $Log(A)$ is the set of all strings in $\Sigma_\infty$ such that inclusion in any word in $\Sigma^*$ of a string from $Log(A)$ is sufficient to ensure that that word belongs to cylindrification of A.
	
Then $Log(Exp(H))$ is the set of all strings in $\Sigma_\infty$ whose presence in a word from $\Sigma^*$ would be a guarantee for that word to belong in the cylindrification of $Exp(H)$. However, the cylindrification of $Exp(H)$, noted $Exp(Exp(H))$, reduces to just $Exp(H)$. Then, $Log(Exp(H))$ is the set of all strings in $\Sigma_\infty$  whose presence in a word from $\Sigma^*$ would be a guarantee for that word to belong in $Exp(H)$.

But then $Log(Exp(H))$ is the set of all strings in $\Sigma_\infty$  whose presence in a word $x \in \Sigma^*$ serve as symptom that word $x$ also includes a string in $H$.

(I) $H \subseteq Log(Exp(H))$. Indeed, if $x$ includes a string from $H$, this is the most clear symptom that $x$ includes strings from $H$.
	
(II) Let $H \subseteq K$. Let $g$ belong in $Log(Exp(H))$. Let $x$ include $g$, then it also includes a string $f$ from $H$. But $f \in K$, then $g$ is also a symptom for presence in $x$ of a string from $K$. Hence $g \in Log(Exp(H))$.
	
(III) $LogExp(H)=LogExp(LogExp(H))$. This is actually obvious.
\end{proof}

We may call just $LogExp$ the mapping $\Sigma_\infty \rightarrow \Sigma_\infty$  which carries $H$ onto $Log(Exp(H))$. Thus, $LogExp$ is a closure operator. This closure is not topological. Indeed, it is not always the case that $LogExp(H \cup K)=LogExp(H) \cup LogExp(K)$. There are cases where $LogExp(H \cup K)$ contains certain special kinds of a string $g$ whose presence in a word behaves as a spy for the presence in that word of a string which in some cases is from $H$ and in others is from $K$. Clearly, such a string $g$ couldn't be a member of either $LogExp(H)$ or $LogExp(K)$. String $g$ would serve as a collective sympton for both strings from $H$ and from $K$. This is a key point in this research.
\begin{theorem}
Given $A \subseteq E$, $Exp_E (Log_E (A))=E^A$.
\end{theorem}
\begin{proof}
(I) Let $x \in Exp_E (Log_E (A))$. Then $x$ is a word in $E$ that includes at least a string $g \in Log_E (A)$. By definition of logogram, $x$ is in $E^A$.

(II) Let $x \in E^A$. Thus, $x$ is a word in $E$ which includes a string from $A$.  Since $A \subseteq Log_E (A)$, we also have that $x$ is a word in $E$ which includes a string from $Log_E (A)$. The set of all words in $E$ which include strings from $Log_E (A)$ is $Exp_E (Log_E (A))$. Hence $x \in Exp_E (Log_E (A))$.
\end{proof}

Note that, since $Log_E (A)$ and $|Log_E (A)|$ are isoexpansive, we also have $Exp_E|Log_E (A)|=E^A$.

\section{Kernel of a Decision Program}
With any NP problem $(E, F)$ we associate a set of strings $|Log_E(F)|$ called the reduced logogram of $F$ relative to $E$, which conveys structural information on $E$, $F$, and how $F$ is embedded in $E$. We assume $F$ to be a relative cylinder in $E$ (the rationale for this assumption will be given in Section 12).

The strings in $|Log_E (F)|$ serve as certificates of membership for $F$ relative to $E$. This means that, limited to words in $E$, to include one or more strings from $|Log_E (F)|$ is necessary and sufficient for membership of a word in $F$.

In principle, we cannot exclude that $|Log_E (F)|$ may contain strings that behave as collective witnesses, also called wizards. There exist problems, e.g. $PRIMES$, where $|Log_E (F)|$ has wizards. (Note, incidentally, that $PRIMES$ is in P \cite{agrawal}.)
	Should that be the case, a program $P$ solving $(E, F)$ might do calculations that are functionally equivalent to testing input $x$ for wizards: That would make the task of deciding about input easier.

Let program $P$ solve problem $(E, F)$. The tests in $|Log_E (F)|$ are those that $P$ can use: They are so to speak at disposal for a program $P$. Which of these tests are actually used by $P$ is a different story. We define the $kernel$ of program $P$, noted $Ker(P)$, to be the set of the strings from $|Log_E (F)|$ that $P$ actually uses for making decisions. The strings in $Ker(P)$ are uniquely identified by the algorithm that $P$ implements. The composition of $Ker(P)$ in terms of strings can also be determined through experiments with the executable of $P$.

Let $H$ be a subset of the reduced logogram $|Log_E (F)|$ of $F$ relative to $E$. We call $H$ complete for problem $(E, F)$ as soon as, for any $x \in E$, we have that $x$ actually belongs to $F$ if and only if $x$ includes at least one string $g$ from $H$. If no proper subset of the reduced logogram $|Log_E (F)|$ happens to be complete for $(E, F)$ then we say that $|Log_E (F)|$ is irreducible. Since $Ker(P)$ is a subset of the reduced logogram, when this last happens to be irreducible we have that any two programs that solve $(E, F)$ have same program kernel, written $Ker(P)=Ker(Q)$. This by no means implies that $P$ and $Q$ shall have to exhibit equal time complexities, however we can make informal remarks that suggest that further investigation of the matter might be worthwhile in this case.

In companion paper we proved that, for $E=CNF, F=SAT$, problem $(E, F)$ cannot have collective certificates in its reduced logogram, and this in turn yielded that its reduced kernel $|Log_E (F)|$ is irreducible. We derived these results from a property of structural independence of SAT that we called ``strong internal independence."

In what follows we prove that SAT exhibits a strictly stronger form of structural independence, that we call ``complete internal independence." We argue that this form of independence of a decision problem is the strongest possible for a decision problem. Using this stronger property, we will be able to give more explicit form to the informal remarks that we made in \cite{dizenzoarxiv} on the time complexity of SAT.

\section{Complete Independence of events}
We consider a decision problem $(E, F)$. We already noted that, if $g \in |Log_E (F)|$, then $Exp_E(g)=Exp(g) \cap E$ is fully included in $F$. Thus, $Exp_E (g)$ is an elementary cylinder relative to $E$, and is fully included in $F$. The target $F$ of problem $(E, F)$ is covered with sets $Exp_E (g)$ where $g \in |Log_E (F)|$. Statements such as $x \in Exp_E (g)$ with $g \in |Log_E (F)|$ will be called ``events in the universe $F$," and we can rely upon the fact that both the union and the intersection of two events in universe $F$ are events in universe $F$.

Here the events in the universe $F$ are subsets of $F$ being closed in $E$. We are using ``event" for subsets of $E$ of the form $E^H$ where $H \subseteq \Sigma_{\infty} (E)$.

Thus, given any decision problem $(E, F)$, we introduce the $cover$ of the target set $F$ associated with $|Log_E (F)|$ to be the family of sets
\begin{equation}
\mathcal D_E (F) = \{Exp_E (g) \subseteq F : g \in |Log_E (F)|\}.
\end{equation}
Its members are the $regions$ of the cover. The cover that is associated with the kernel of a program $P$ solving $(E, F)$ is then
\begin{equation}
\mathcal F_P (E, F) = \{Exp_E (g) \subseteq F : g \in Ker(P)\}.
\end{equation}
Both $\mathcal D_E (F)$ and $\mathcal F_P (E, F)$ are collections of subsets of the target set $F$ whose union is $F$, with $\mathcal F_P (E, F)$ being a subcollection of $\mathcal D_E (F)$.

We are now in good position to define a notion of complete independence of a finite set of events in the universe $F$. By a partition of the set $F$ we understand a collection of events within universe $F$ being pairwise incompatible and exhaustive. Two events in universe $F$ are incompatible as soon as the corresponding sets are disjoint. By a collection of events in universe $F$ being exhaustive we understand that, if input $x$ is in $F$, then at least one of them will occur.

Consider a finite set of events $E_1, E_2, .., E_m$. We will define a notion of complete independence among them. We will reduce the general case to the simple case of a partition. To this end, we take under consideration the $2^m$ products $U_1U_2 .. U_m$ where $U_i$ can be either $E_i$ or its complement $E_i^c = F-E_i$. We may obtain the $2^m$ formal products by developing the form
\begin{displaymath}
(E_1+E_1^c)(E_2+E_2^c)..(E_m+E_m^c)
\end{displaymath}
Here product means intersection and we use sum for union to stress that it is disjoint. Some of the $2^m$ products can be void, and we do not take care of them. Those that are nonvoid take the name of atomic constituents $C_1, C_2,.., C_s$ of the partition induced by events $E_1, E_2, .., E_m$ where $s \le 2^m$.

We say that $m$ events $E_1, E_2, .., E_m$ are completely independent as soon as they give rise to $2^m$ nonvoid atomic constituents. If $m$ events are completely independent, then every one of them remains uncertain (we do not know if it happened) even if we are notified the outcome of each of the other $m-1$.

\section{Witnesses and Wizards}
The computations that a decision program $P$ performs on an input $x$ are functionally equivalent to sequences of tests done on $x$. The tests that $P$ can perform on an input, hence those that can occur in one such sequence of tests, are those that search the input word for strings $g \in Ker(P)$. This is a straight consequence of how we defined $Ker(P)$. We actually assume that $P$ can only do calculations that encode tests on $x$ that belong to this collection of tests.

Searching $x$ for a string $g$ amounts to asking if $x$ happens to belong in the absolute elementary cylinder $Exp(g)$ associated with $g$. We thus arrive at the conclusion that all that $P$ can possibly do to arrive at a decision consists in asking questions of this form. Note that $P$ has not got to ask whether $x$ is in $Exp_E (g)$ since $P$ already knows that $x$ is in $E$. This is an important point since asking if $x \in Exp_E (g)$ would be more computationally expensive. Thus, as long as we are dealing with words whose membership in $E$ has already been ascertained, we can freely exchange the relativized cylinders $Exp_E (g)$, $Exp_E (f)$,.. with the corresponding absolute cylinders $Exp(g)$, $Exp(f)$,.. . \vspace{0.3cm}

In our theory, the state of knowledge of a running program $P$ at any stage during computation $P(x)$ on input $x$ consists of a pile of assertions, namely those that have been collected up to that computation stage. The assertions that are derived in one and same computation are certainly mutually consistent (since they assert properties pertaining to one and same object $x$). We are indebted to Dana Scott for this style of looking at computations. When, for some $g \in Ker(P)$, program $P$ asks if $x$ belongs in $Exp(g)$ (more pedantically, when $P$ performs calculations that amount to testing if $x$ belongs in $Exp(g)$) this question always gets an answer. The answer can be a ``yes" when the test is passed, or a ``no" when it is failed. In case of a negative answer, the new piece of information ``$x$ is not in $Exp(g)$" is acquired on part of the program $P$. Since $Exp_E (g)$ is a subset of $Exp(g)$, the more interesting piece of information ``$x$ is not in $Exp_E (g)$" can be inferred from ``$x$ is not in $Exp(g)$." Thus, in this case the state of knowledge of program $P$ gets enlarged by the addition of the new piece of information ``$x$ is not in $Exp_E (g)$."  In case of success the new piece of information which is acquired is of course ``$x$ is in $Exp_E (g)$." Since it is already known to $P$ that $x$ is in $E$, in this case the state of knowledge of program $P$ gets enlarged by the addition of the new piece of information ``$x$ is in $Exp_E (g)$." (Remarks on what program knows at various stages in a computation shall not be regarded as  ventursome as they can be made formal using methods of model theoretic analysis of program knowledge \cite{fagin}.) \vspace{0.3cm}

In this theory, information regarding $x$ is acquired by $P$ in lumps. The acquisition of a piece of information occurs at the moment when the execution of a sequence of tests is completed. We may well think of a piece of information as being a piece of paper carrying a written note such as ``$x$ is in $Exp_E (g)$" or else ``$x$ fails to be in $Exp_E (g)$." These notes stack one upon the other until the pile becomes a decisive one: This is the case when the data that was gathered entails one of the events $x \in F$ or $x \in E-F$. (It must be given credit to Scott for these conceptual contents of the theory.)

Since $Ker(P)$ is a subset of $|Log_E (F)|$ which is a reduced set of strings, it cannot be that $Ker(P)$ contains two substrings $f, g$ with $f$ included in $g$. Hence the cylinder associated with, say, $f$ cannot include the cylinder associated with $g$. Thus, $P$ never tests membership of input $x$ into two cylinders one included in the other.

It is instead possible that a cylinder $Exp_E(g)$ with $g$ in $Ker(P)$ will intersect one or more other such cylinders, and be completely included in the union of the intersected cylinders. This can happen when $g$ is a witness for a whole bundle of solutions, or (in somewhat more esoteric manner) when $g$ is a wizard. The next two paragraphs have the details.

\paragraph{Witnesses and wizards formally defined}
Assume that we are considering an input word $x$ of size $n$. Let $g$ be a string in the reduced logogram $|Log_E (F)|$ of an NP problem $(E, F)$ so that $Exp_E (g)$ is a subset of the target set $F$. We assume that the size of string $g$ is less or equal to $n$, which is written $|g| \le n$, thus $g$ actually belongs in $|Log_E (F^n)|$.
	
Let us remember that, given any NP problem $(E, F)$, there is a decomposition of the target set $F$ into subsets $F_i$ called solution regions, where $F_i$ is the set of those words $x$ in $E$ that are satisfied by solution $y_i$.
	
That being granted, we say that $g$ is a witness as soon as its associated relativized cylinder $Exp_E (g)$ is fully included in at least one of the $F_i$s. We say that $g$ is a proper witness as soon as its associated relativized cylinder $Exp_E (g)$ is fully included in exactly one of the $F_i$s. Equivalently, $g$ is a proper witness as soon as $g$ is an encoded sign of satisfiability that points toward a unique solution $y_k$. If $Exp_E (g)$ is included in the intersection of two or more of the $F_i$s then $g$ is an improper witness (also called a pseudowizard).
	
We say that $g$ is a wizard as soon as $Exp_E (g)$ fails to be fully included in a solution region. Thus, when $g$ is a wizard, its inclusion in a word $x$ belonging to the reference set $E$ is a guarantee that $x$ is satisfiable while nothing can be said about what particular solutions satisfy $x$.

\paragraph{Immediate properties of the wizards}
Take $g \in |Log_E (F)|$, and assume that $Exp_E (g)$ fails to be fully included in any one single solution region $F_i$. Then $g$ is a wizard. To simplify things, assume that $Exp_E (g)$ is fully included in the union of two solution regions $F_h$ and $F_k$. Then the situation is
\begin{equation}
Exp_E(g) \subseteq F_h \cup F_k, Exp_E(g) \not\subseteq F_h, Exp_E(g) \not\subseteq F_k.
\end{equation}
Let us develop our remarks for some fixed echelon $E^n$, where $n \ge |g|$. Let $f_1,.., f_s$ be all the strings in $\cup_{i=1}^{\alpha(n)} |Log_E (F_i^n)|$ whose associated relativized cylinders intersect $Exp_E (g)$:
\begin{equation}
Exp_E(g) \cap Exp_E(f_j) \not= \emptyset, j=1,.., s
\end{equation}
The reader will remember that $\cup_{i=1}^{\alpha(n)} |Log_E (F_i^n)|$ is the set of all the witnesses that exist in the reduced logogram $|Log_E (F^n)|$. Thus, the strings $f_1,.., f_s$ are exactly those witnesses in the reduced logogram $|Log_E (F^n)|$ whose associated relativized cylinders happen to intersect $Exp_E (g)$.

The strings $f_1,.., f_s$ will be referred to as the witnesses associated with $g$.

We set $Exp_E(f_j) = C_j$ all $j = 1,.., s$. Note that none of sets $C_1,.., C_s$ is included in $Exp_E(g)$. Indeed, should $C_k$ be a subset of $Exp_E(g)$, we would have $g \le f_k$, an absurd since both $g$ and $f_k$ are members of the reduced set  $|Log_E (F^n)|$.

We conclude this section proving a theorem.

\begin{theorem}
If $g$ is a wizard and $f_1,.., f_s$ are its associated witnesses, then $Exp_E(g)$ is properly included in the union of the corresponding cylinders $C_1 = Exp_E(f_1),.., C_s = Exp_E(f_s)$.
\end{theorem}
\begin{proof}
Let $(E, F)$ be any NP problem with $F_1,.., F_{\alpha(n)}$ as solution regions associated with $n$th echelon $(E^n, F^n)$.

It follows from Theorem 7 of \cite{dizenzoarxiv} that the set $\bigcup_{i=1}^{\alpha(n)} |Log_E (F_i^n)|$ of all witnesses in the reduced logogram $|Log_E (F^n)|$ is complete for problem $(E, F)$.

This implies that the union of the relativized cylinders associated with the strings in $\bigcup_{i=1}^{\alpha(n)} |Log_E (F_i^n)|$ is coincident with the whole of the target set $F^n$ of problem $(E^n, F^n)$. (In this proof we are reasoning echelon by echelon.)

As consequence, the relativized cylinder $Exp_E(g)$ is coincident with the union of its intersections with cylinders $C_1,.., C_s$.

Hence, $Exp_E(g) \subseteq C_1 \cup C_2 \cup .. \cup C_s$.

By contradiction, let us assume that $Exp_E(g) = C_1 \cup C_2 \cup .. \cup C_s$.

Then $C_1 \cup C_2 \cup .. \cup C_s$ is an elementary relativized cylinder, and $g$ is its signature.

From $Exp_E(g) = C_1 \cup C_2 \cup .. \cup C_s$ follows $C_j \subseteq Exp_E(g)$ all $j=1,.., s$.

$C_j \subseteq Exp_E(g)$ is rewritten $E \cap Exp(f_j) \subseteq E \cap Exp(g)$, and from this last inclusion follows $Exp(f_j) \subseteq Exp(g)$ all $j=1,.., s$.

Hence, $g \le f_j$ all $j=1,.., s$. Absurd, since $g$ and $f_j$ are members of $|Log_E (F^n)|$, which is a reduced set of strings.
\end{proof}

We conclude that, if $g$ is a wizard, then, with notations given, (i) $Exp_E(g)$ is a proper subset of $C_1 \cup .. \cup C_s$, and (ii) $C_j \not\subseteq Exp_E(g)$ and $Exp_E(g) \not\subseteq C_j$ all $j=1,.., s$.

Note that (i)-(iii) is not sufficient to conclude that $g$ is a wizard.

\section{Structural Independence of Problems}
One of the themes of this paper is the study of finite collections $\mathcal F$ of subsets of the target $F$ of a decision problem $(E, F)$. This study is related with our interest for notions of structural independence of decision problems.

We have seen that both $\mathcal D_E (F)$ and $\mathcal F_P (E, F)$ are collections of subsets of the target set $F$ of problem $(E, F)$, with $\mathcal F_P (E, F)$ being a subcollection of $\mathcal D_E (F)$. We also have seen that the elements of these collections are elementary relativized cylinder sets. We defined $\mathcal D_E (F)$ to be the collection of all sets of the form $Exp_E (g)$ where $g \in |Log_E (F)|$: Thus, there is one-one onto correspondence between the elementary relativized cylinders in $\mathcal D_E (F)$ and the strings in $|Log_E (F)|$. $\mathcal F_P (E, F)$ is the subcollection of $\mathcal D_E (F)$ that is associated with $Ker(P)$.

\paragraph{Pairwise independence of strings}
 Let $f, g \in \Sigma_{\infty} (E)$ for the whole paragraph. We say that $f$ entangles $g$ relative to $E$ as soon as all $x \in E$ which include $f$ also include $g$ (note that since $f, g \in \Sigma_{\infty} (E)$ there exists $x \in E$ which includes $f$ and there exists $y \in E$ which includes $g$). In \cite{dizenzoarxiv} we developed a theory of entanglement among strings. That $f$ entangles $g$ relative to $E$ was denoted $f \sqsupseteq^E g$.

$f, g$ independent relative to $E$ means that neither $f \sqsupseteq^E g$ nor $g \sqsupseteq^E f$.
Thus, $f, g$ independent relative to $E$ means that (i) there is $x \in E$ which includes $f$ and does not include $g$ and (ii) there is $y \in E$ which includes $g$ and does not include $f$.

Note that $f \sqsupseteq^E g$ if and only if $Exp_E (f) \subseteq Exp_E (g)$. Analogously for $g \sqsupseteq^E f$. Then $f, g$ independent relative to $E$ means that neither $Exp_E (f) \subseteq Exp_E (g)$ nor $Exp_E (g) \subseteq Exp_E (f)$.

\paragraph{Internal independence of a problem}
We call problem $(E, F)$ internally independent as soon as the strings in $|Log_E (F)|$ are pairwise independent relative to $E$.
Thus, $(E, F)$ internally independent $\equiv$ $\mathcal D_E (F)$ is an antichain.

\paragraph{Strong internal independence}
In our previous paper \cite{dizenzoarxiv} we defined the property of strong internal independence of a decision problem. We defined problem $(E, F)$ to have this property as soon as for any choice of $s$ distinct strings $f_1,.., f_s$ in $|Log_E (F)|$, the following is true: For every $i$ between $1$ and $s$ there exists a word $x_i \in E$ such that $x_i$ contains $f_i$ and fails to contain any of the remaining strings in $\{f_1,.., f_s\}$.
\begin{theorem}	
Strong internal independence of a decision problem implies internal independence.
\end{theorem}
\begin{proof}
By contradiction, assume that $(E, F)$ exhibits the strong internal independence property and does not exhibit the simple dependence property.

Thus, not all of the strings in $|Log_E (F)|$ are independent relative to $E$ taken two by two.
This is to say that there exist $f, g \in |Log_E (F)|$ such that $Exp_E (f) \subseteq Exp_E (g)$.

With reference to the definition of the strong internal independence property we take $f, g$ as a particular choice for $f_1,.., f_s$. Since $(E, F)$ has the strong internal independence property we have that (i) there exists $x \in E$ such that $x \ge f$ and $x \not\ge g$, (ii) there exists $y \in E$ such that $y \ge g$ and $y \not\ge f$.

On the other side, since $x \ge f$ and $x \in E$, we have that $x \in Exp_E (f)$. By the contradiction hypothesis, we also have $x \in Exp_E (g)$. This implies $x \ge g$, an absurd.
\end{proof}

\paragraph{Complete internal independence}
We shall say that problem $(E, F)$ has the complete internal independence property as soon as, given any finite set $f_1,.., f_s$ of pairwise compatible strings in $|Log_E (F)|$, there exists $x \in E$ such that (i) $x$ includes each of the $f_1,.., f_s$, (ii) $x$ fails to include any other $g \in |Log_E (F)|$ except possibly those that are subsumed by $f_1+..+f_s$.

In Section 9 we defined complete independence of a finite collection of sets in $\mathcal D_E (F)$. We may well rephrase our definition of complete independence of a decision problem as follows: $(E, F)$ is completely independent if any finite collection $E_1,.., E_s$ of pairwise intersecting regions in $\mathcal D_E (F)$ is completely independent.

\section{Application to SAT}
The encoding scheme that we adopt converts $CNF$ formulas into words over $\Sigma=\{0, 1, 2\}$. In what follows $E=CNF$, $F=SAT$.
	
We represent clauses over $x_1,.., x_n$ by sequences of $n$ codes from $\Sigma$. Code $0$ denotes absence of the variable, code $1$ presence without minus, code $2$ presence with minus.
E.g., clause $x_1 \vee x_3 \vee -x_4$ becomes 1012.

A whole formula is encoded as a sequence of clauses. We define $F^{nm}$ = satisfiable formulas with $n$ variables and $m$ clauses.

Every encoded formula has a prefix of the form 0..010..01 consisting of $n$ 0s followed by a 1 followed by $m$ 0s followed by a 1. A program $P$ solving $(E, F)$ learns the current values of $n, m$ from this prefix. $P$ shall have to be aware that the $nm$ characters on the immediate right of the prefix encode $m$ clauses over $n$ boolean variables.

The reference set $E=CNF$ is thus encoded as a prefix-free language over $\Sigma=\{0, 1, 2\}$. Given any subset $A \subseteq E$, one has $Exp_E (A)=Exp(A) \cap E=A$. Thus, any subset $A \subseteq E$ is a cylinder relative to reference set $E$. This is to say that any $A \subseteq E$ is a closed set in $E$. Note that these simplifications fairly match with usual programming practice. Note that the target $F$ of problem $(E, F)$ is in any case a cylinder relative to $E$, as required by theory developed above.\vspace{.2cm}

\noindent
The $size$ (or $complexity$) $|x|$ of a boolean formula $x$ is the number of distinct variables that have occurrences in $x$. By the $effective$ $size$ of a satisfiable boolean formula $x$ we understand the minimum number of value assignments to variables that are needed to evaluate the formula to 1. We define the effective size of an unsatisfiable formula to be the size of the formula.

$Example$ Consider formula $x = (x_1 \vee x_3 \vee -x_4) \wedge (x_2 \vee -x_3)$. The size is 4, the effective size is 2. Indeed, the two value assignments to variables $x_1 = 1$ and $x_2 = 1$ are sufficient to set $x$ to 1. This is not the only partial value assignment to variables that satisfies $x$. For example, the partial assignment $x_4=0, x_3=0$ is also sufficient to set the value of $x$ to 1.

If size and effective size of a formula are different we say the formula is $bewitched$ (a formula is $unbewitched$ if the two numbers are equal). If $x$ is bewitched, then certainly $x$ includes a pseudowizard. Let us verify this on formula $x$ in the above example: The encoded version of $x$ is
$000010011\flat12\flat12\flat$. It subsumes the strings $000010011\flat\flat\flat\flat1\flat\flat$ and
$00001001\flat\flat\flat2\flat\flat2\flat$ that are both pseudowizards. (Remember that pseudowizards are witnesses.)

The unbewitched formulas form a hardest subset of $SAT$, and we may ignore bewitched formulas without loss of generality. \vspace{.2cm}

\noindent
We introduce the sequence $y_1, y_2, ..$ of solutions, and the corresponding sequence $F_1, F_2, ..$ of recursive subsets of $F$. Here the solutions $y_i$ consist of value assignments. The cardinality function is  $\alpha(n)=2^n$.
	The target set $F=SAT$ as well as the solution regions $F_1, F_2, ..$ are closed sets in $E=CNF$. Thus, all these sets are relative cylinders in $E$. These assumptions correspond to properties of $SAT$ that can be derived under various other encoding schemes \cite{balcazar} \cite{hemaspaandra}.

Before we prove our main result in this paper, let us spend a few words on the logogram of $SAT$. A string in $|Log_E (F^{nm})|$ is a prescription that a word in $F^{nm}$ may or may not be conformant with. We may represent a string in $|Log_E (F^{nm})|$ as a word of length $nm$ over $\{\flat \}\cup \Sigma$ (ignoring prefix).
	Example for $n=m=3$: String $\flat \flat 11 \flat 2 \flat 2 \flat$  prescribes that first clause shall include $x_3$, second shall include $x_1$ and $-x_3$, third shall include $-x_2$. Note that strings in $|Log_E (F^{nm})|$ only prescribe either $1$ or $2$ as values (by the minimality property of reduced logogram).

\begin{theorem}
Let $E=CNF, F=SAT$. Problem $(E, F)$ exhibits the complete internal independence property.
\end{theorem}

\noindent
$Outline$ $of$ $proof$. We prove that, given $s$ distinct, pairwise compatible strings $f_1,.., f_s$ from the reduced logogram $|Log_E (F^{nm})|$, there is $x \in E$ such that (i) $x \ge f_1+f_2+..+f_s$, (ii) for all of the remaining $g \in |Log_E (F^{nm})|$ one has $x \ge g$ if and only if $g \le f_1+f_2+..+f_s$.

\begin{proof}
We consider $s$ distinct strings $f_1,.., f_s$ taken from $|Log_E (F^{nm})|$. Thus, regarded as a partial function, each $f_i$ will assign only values $1$ or $2$. We assume that $f_1,.., f_s$ are compatible taken two by two.

We shall rely upon the main result proven in \cite{dizenzoarxiv} namely that SAT has no wizards. By virtue of this result, none of the strings $f_1,.., f_s$ is a wizard. That means that each of them is a witness, which in turn means that each of them consists of a consistent prescription of exactly one literal to each of the $m$ clauses. That a prescription is consistent means that it does not comprise (i) an assignment of a literal to a clause and (ii) the assignment of the negative of that literal to another clause.

Let us consider any two of the strings $f_1,.., f_s$, call them $f_h$ and $f_k$.

Since $f_h$ and $f_k$ are compatible, we cannot find two distinct integers $i, j$, $1 \le i \le n, 1 \le j \le m$, such that (a) $f_h$ assigns literal $x_i$ to clause $j$, (b)  $f_k$ assigns literal $-x_i$ to same clause $j$.

Since that holds for any pair $f_h, f_k$, we take it for granted that, if one of the strings $f_1,.., f_s$ prescribes a literal to one of the clauses, none of the remaining strings within $f_1,.., f_s$ will prescribe the negative of that literal to the same clause.

That being granted, we define $x$ to be that uniquely identified formula that has a literal in a clause if and only if at least one of the strings $f_1,.., f_s$ prescribes that literal to that clause.

Formula $x$ subsumes everyone of the strings $f_1,.., f_s$, hence $x$ subsumes their join $f_1+f_2+..+f_s$.

Let $g \in |Log_E (F^{nm})|$ and assume $x \ge g$.

By way of contradiction, assume that $f_1+..+f_s \not\ge g$. We write $f$ for $f_1+..+f_s$ so that we have $Dom(f)=Dom(f_1) \cup..\cup Dom(f_s)$. The hypothesis taken by contradiction is rewritten $f \not\ge g$.

From our definitions about strings, we have that $f \ge g$ if and only if $Dom(g) \subseteq Dom(f)$ and, for any $i \in Dom(g)$, $f(i)=g(i)$. Then, since $f \not\ge g$, only the following two cases are possible:

Case 1. $Dom(g) \not\subseteq Dom(f)$.

In this case, there exists $k \in Dom(g)$ such that $k \not\in Dom(f)$. This last is rewritten $k \not\in Dom(f_1),.., k \not\in Dom(f_s)$.

We made the assumption $x \ge g$. Since $g$ prescribes only $1$ or $2$ as values, this assumption implies $x(i) \not= 0$ for all $i \in Dom(g)$. In particular, we have $x(k) \not= 0$. This is absurd since, by the above construction of formula $x$, we have $x \not= 0$ only in $Dom(f)$.

Case 2. $Dom(g) \subseteq Dom(f)$ but there is an argument value $i \in Dom(g)$ where $f(i) \not= g(i)$.

Since $Dom(g) \subseteq Dom(f)$, we certainly have $i \in Dom(f_1)$ or $i \in Dom(f_2)$ or.. or $i \in Dom(f_s)$. With no loss of generality, we assume $i \in Dom(f_1)$.

We then have $g(i) \not= f_1(i)$.

However, since $x \ge g$ and $i \in Dom(g)$, we also have $x(i)=g(i)$. Then $x(i) \not = f_1(i)$, a contradiction since $f = f_1$ everywhere in $Dom(f_1)$.
\end{proof}

\section{On the Time Complexity of SAT}
In this section we set forth a hint of proof of the exponentiality of SAT in the light of Theorem 10 above together with Theorems 8, 9, 10 of companion paper. Remember that $E=CNF$, $F=SAT$. We conjecture that a formal proof can be derived following this hint.
\begin{theorem}
SAT is exponential.
\end{theorem}

\noindent
Our hint of proof consists of two parts, (I) and (II). \vspace{.3cm}

\noindent
(I) It follows from Theorem 10 of \cite{dizenzoarxiv} that there is a unique subfamily $\mathcal F$ of $\mathcal D_E (F)$ such that $F = \bigcup \mathcal F$, namely $\mathcal F = \mathcal D_E (F)$ itself. As a consequence, for any proper subfamily  $\mathcal F \subset \mathcal D_E (F)$ one has $F \not= \bigcup\mathcal F$.

We then have that it cannot be that $\mathcal F_P (E, F)$ is a proper subfamily of the full cover $\mathcal D_E (F)$, otherwise we would have $F \not= \bigcup\mathcal F_P (E, F)$, and then $P$ could not be correct as a program. In particular, since the cardinality of $\mathcal D_E (F)$ grows exponentially with word size, we have that $\mathcal F_P (E, F)$ is not allowed to be a polynomial subfamily of $\mathcal D_E (F)$. Thus, no search algorithm for $SAT$ can only search a polynomial family of sets.\vspace{.3cm}

\noindent
(II) It remains for us to discuss the possibility that one single algorithm can solve the full search problem for $x$ by directly searching the full exponential family $\mathcal D_E (F)$ in polynomial time. However this can scarcely be the case due to complete absence of any form of dependence among subsets in the reduced logogram $|Log_E (F)|$ for $E=CNF$, $F=SAT$. By this lack of internal dependence, any computation of a program $P$ solving $(E, F)$ is such that the result of any computation step does not change the results that are left possible for the subsequent steps. In the rest of this part we make a few informal remarks on how this lack of dependence comes into play.

We take a general purpose program machine $M$ as computation model. (That $M$ is a program machine means that the process carried out by $M$ is determined by a running program.) We assume that only one program is running at any moment of time within $M$. We keep machine $M$ fixed while we consider an infinite set of programs solving $SAT$ (actually the set of all programs that run on $M$ and solve $SAT$). We emphasize that the hardware is kept fixed while different programs all running on that hardware are compared.

Let $B(x, k)$ be a program running on machine $M$ which for any given input $x \in E$ of size $n$ and every integer $k$ between $1$ and $2^n$ will decide if $x$ has solutions in the range between $y_1$ and $y_k$. Take $Time_B(x, k)$ be the number of time units that algorithm $B$ uses on inputs $x, k$ on machine $M$. \vspace{.2cm}

(II$\alpha$) We consider $\alpha(n)=2^n$ strings from reduced logogram $|Log_E(F^{nm})|$ call them $f_1,.., f_{\alpha(n)}$. Since $SAT$ has no wizards, each of these strings is a witness. By way of consequence, each $f$ within $f_1,.., f_{\alpha(n)}$ is associated with one bundle of solutions (one single solution when $f$ is a proper witness).

With no loss of generality, we may choose the strings $f_1,.., f_{\alpha(n)}$ to be all proper witnesses. Besides, we may well choose the strings so that, for $j=1,.., \alpha(n)$, the unique solution associated with $f_j$ is $y_j$. Note that we do not assume that the strings $f_1,.., f_{\alpha(n)}$ are pairwise compatible.

By the strong internal independence of $SAT$, for every $j=1,.., \alpha(n)$ there exists a word $x \in F^{nm}$ which includes string $f_j$ and fails to include all of the remaining strings in $f_1,.., f_{\alpha(n)}$. Thus, $x$ is satisfied by solution $y_j$. Nothing can be said of the remaining solutions since the fact that $x$ fails to include $f_i$ where $i\not=j$ does not exclude that $x$ can possibly include other witnesses associated with solution $y_i$.

However, in the above argument we may keep $f_j$ fixed while varying, for all $i=1,..,\alpha(n), i\not=j$, the string $f_i$ over $|Log_E(F_i^{nm})|$ in any possible manner. We then conclude that, for any single $j$ between $1$ and $2^n$, there exists an $x \in F^{nm}$ which is satisfied by $y_j$ and fails to be satisfied by any other solutions $y_i \not= y_j$.

In particular, given any integer $k$ such that $1 \le k < 2^n$, there exists $x \in F^{nm}$ which is satisfied by solution $y_{k+1}$ but is not satisfied by any one of the solutions in the range from $y_1$ and $y_k$. Thus, it may well be the case that $B(x, k)=0$ and $B(x, k+1)=1$. This statement is a prerequisite for what follows.

(The above argument may seem to us very obvious in the light of our empirical understanding of $SAT$. The internal independence properties of this problem put this on deductive bases.)
\vspace{.3cm}

(II$\beta$) We will prove the following.

If for any unsatisfiable $x \in E^{nm}$ and any $k<2^n$ we have $Time_B (x, k) = Time_B (x, k+1)$, then there exists another program $A_x$ solving $SAT$ on hardware $M$ such that
\begin{equation}
Time_{A_x} (x, k) < Time_{A_x} (x, k+1),
\end{equation}
\begin{equation}
Time_{A_x} (x, 2^n) \le Time_B (x, 2^n).
\end{equation}
Indeed, under the above hypotheses on $M$, we can speak of the class of all programs $B$, $C$,.. that solve $SAT$ on machine $M$. Given any input $y$ of size $n$, we can then introduce a most efficient program $A_y$ for particular input $y$ in this class. We understand that $A_y$ is a most efficient program for a particular input $y$ as soon as $A_y$ solves $SAT$ on machine $M$ and, besides, $Time_{A_y}(y, 2^n) \le Time_C(y, 2^n)$ for any other program $C$ solving $SAT$ on hardware $M$. Thus $A_y$ is no worse than any other $C$ solving $SAT$ on $M$ limited to this particular input $y$.

We shall prove that, for an unsatisfiable input $x$, we have $Time_{A_x}(x, i) < Time_{A_x}(x, i+1)$ all $i$, $1 \le i < 2^n$.

If we ignore improper witnesses (which is to say that we disregard bewitched formulas) we have:

\begin{equation}
|Log_E(F_{i+1})| \cap\bigcup_{j=1}^i |Log_E(F_j)| = \emptyset
\end{equation}

From Theorem 10 of present paper follows
\begin{equation}
\bigcup_{j=1}^i |Log_E(F_j)| \not\sqsupseteq^E |Log_E(F_{i+1})|
\end{equation}
\begin{equation}
|Log_E(F_{i+1})| \not\sqsupseteq^E \bigcup_{j=1}^i |Log_E(F_j)|
\end{equation}
(For brevity, in this informal hint we omit the proof that Equations 18, 19 follow from the complete independence of $SAT$.)

Thus, the two sets of strings $|Log_E(F_{i+1})|$ and $\bigcup_{j=1}^i |Log_E(F_j)|$ are disjoint and there is no entanglement between them.

Since $A_x$ is optimal on input $x$, algorithm $A_x (x, i)$ may perform on $x$ calculations that implement the search of word $x$ for strings in $|Log_E (F_{i+1})|$ only if these calculations can be performed within the same set of machine cycles that are allocated to the calculations that implement the search of word $x$ for strings in $\bigcup_{j=1}^i |Log_E(F_j)|$.

For this overlap to be an affordable statement it is required that the first search (the one for strings in $|Log_E (F_{i+1})|$) should be implemented by the same set of calculations that implement the second search (the one for strings in $\bigcup_{j=1}^i |Log_E(F_j)|$) so that, while performing the second search, also the first would be implicitly executed. By theory developed in \cite{dizenzoarxiv}, this kind of magics can actually occur through the mechanism of entanglement between sets of strings. That mechanism allows for different computation tasks to be performed by one and same calculation: It is well possible that one and the same computation implements distinct operations of search, acting upon distinct sets of strings. (In present theory computations occur as searches for strings in disguise: The calculations that a decision program $P$ performs actually implement operations of search of an input word $x$ for certain strings in the kernel of $P$.)

 However, from Equations 17-19 it follows that the two sets of strings  $\bigcup_{j=1}^i |Log_E(F_j)|$ and $|Log_E(F_{i+1})|$ are disjoint and there is no entanglement between them. By this lack of entanglement, any overlap between an operation of search for strings in $|Log_E (F_{i+1})|$ and an operation of search in $\bigcup_{j=1}^i |Log_E(F_j)|$ is excluded.

By virtue of (i) this lack of entanglement, together with (ii) the assumed optimality of the search performed by algorithm $A_x (x, i)$ within set $\bigcup_{j=1}^i |Log_E(F_j)|$, we have that the computation that algorithm $A_x (x, i+1)$ performs cannot take less time than the sum of $Time_{A_x} (x, i)$ plus the minimum time needed to search $x$ for strings in $|Log_E (F_{i+1})|$. Since this latter time cannot be $0$, for all those $x \in E^{nm}$ which are not satisfied by solutions in the range from $y_1$ to $y_{i+1}$ we have $Time_{A_x} (x, i) < Time_{A_x} (x, i+1)$. Hence Equation 15 certainly holds for all unsatisfiable $x \in E^{nm}$ and all integers $k$ such that $1 \le k < 2^n$. Equation 16 holds by definition of algorithm $A_x$.

We conclude that an unsatisfiable $x$ requires exponential time on hardware $M$.\vspace{.3cm}

The step from the above informal hint to a possible formal proof is certainly not immediate.
We should at least be able to prove that our model of computation $M$ solves in polynomial time all and only those decision problems that are polynomial on Turing machines. Besides, the theory based on the new model should provide some formal notion of independence between computations.

\section{Conclusions}
We advocated strings as a fundamental notion for studies on computation. Strings are useful to express the notions of simple, strong, and complete internal independence of a decision problem. We have been led to use strings to become able to define the very basic notion of internal independence of a decision problem. Strings seem to be useful since they are absolutely elementary. Strings are already at work in Computability. The ``restrictions" that are often used in the study of circuit complexity are almost one and same notion as strings. (It seems to us that ``string" is the correct name.)

By way of curiosity: Our strings seem to be (from a layman's view) like sampled and discretised versions of the ``strings" that the physicists use in their ``string theories."

Strings are not made of consecutive letters. A string can be interspersed in a word: By canceling zero or more letters in a word $x$, and leaving blanks in places of letters, we get a string  $f$  which is a substring of the original word $x$ (the pun is innocuous: There is no danger in saying that string  $f$  is a substring of $x$). In strings, one has information associated with spaces between letters (and hence with multiple periodicities with which letters may occur in long words).

As soon as we have the strings, we are able to define the kernel of a decision program $P$, noted $Ker(P)$. This is a set of strings which capture structural features of both $P$ and the decision problem $(E, F)$ that $P$ solves.

The program kernel $Ker(P)$ is a subset of $|Log_E (F)|$, the reduced logogram of the target set $F$ in base $E$. The reduced logogram consists of substrings of the words in $F$ which exhibit the following property : If a word in $E$ includes one of these strings then it belongs to $F$. We may think of the strings in $|Log_E (F)|$ as kind of genes of the words in $F$. In early notes the logogram was the $genie$ of problem $(E, F)$. The idea clearly comes from biology, where it is known that certain occurrences at given intervals of certain letters within DNA strings convey structural information, and yield observable characters in the macroscopic development of the structures.
	
Our application to $SAT$ uses structural properties of that problem that seems to have escaped attention so far. We called them ``strong internal independence" and ``complete internal independence." In companion paper we showed that $SAT$ exhibits the strong internal independence property. In that paper we have shown that, by that property, $SAT$ cannot have collective certificates in its reduced logogram. This was our main result in that paper. Starting from that result, we proved in this paper that $SAT$ exhibits a stronger form of structural independence that we called ``complete internal independence."

It seems to us that $SAT$ is difficult due to this extreme form of internal independence: On unsatisfiable inputs, any program solving $SAT$ has exponential worst-case complexity. Our arguments for this conclusion use (i) the complete independence of $SAT$ together with (ii) absence of wizards, that we proved in previous paper. As we are not able, so far, to produce a strictly formal proof (for lack, by far, of a suitable model of computation), we just outlined the ideas of a proof that would use properties (i), (ii). We conjecture that $SAT$ can be proved exponential using the complete independence of $SAT$ together with absence of wizards. We expect that a proof that uses these properties will appear in short time, possibly following the lines of our hint of proof for Theorem 11.

Various relevant candidate proofs of $P \not= NP$ have been set forth in recent years. For some, we do not even have as yet detailed arguments that they fail. A common feature of these efforts is that they try to derive $P \not= NP$ from properties that are known since decades, or use portions of theory (especially one-way functions) that are known since much time. A peculiar feature of our research is that it comes together with a completely new theory, that has applications in diverse fields, and definitely cannot be compressed in few pages. Besides, the property of $SAT$ that is used is completely new in studies on computation.
	
\section{Acknowledgements}
I take the occasion to thank the IBM Semea Director of Research Piero Sguazzero for extremely valuable assistance in the proof checking of \cite{dizenzoarxiv}.

\end{document}